\documentclass[aps, prb, twocolumn, showpacs, superscriptaddress]{revtex4-1}

\usepackage{amsmath,amssymb}
\usepackage{graphicx}
\usepackage{dcolumn}
\usepackage{bm}
\begin{document}

\title {Superfluid density in the slave-boson theory}
\author{Yin Zhong}
\email{zhongy05@hotmail.com}
\affiliation{Center for Interdisciplinary Studies $\&$ Key Laboratory for
Magnetism and Magnetic Materials of the MoE, Lanzhou University, Lanzhou 730000, China}
\author{Han-Tao Lu}
\email{luht@lzu.edu.cn}
\affiliation{Center for Interdisciplinary Studies $\&$ Key Laboratory for
Magnetism and Magnetic Materials of the MoE, Lanzhou University, Lanzhou 730000, China}
\author{Hong-Gang Luo}
\email{luohg@lzu.edu.cn}
\affiliation{Center for Interdisciplinary Studies $\&$ Key Laboratory for
Magnetism and Magnetic Materials of the MoE, Lanzhou University, Lanzhou 730000, China}
\affiliation{Beijing Computational Science Research Center, Beijing 100084, China}

\date{\today}

\begin{abstract}
Despite of the success of the slave-boson theory in capturing qualitative physics of high-temperature superconductors like cuprates, it fails to reproduce the correct temperature-dependent behavior of superfluid density, let alone the independence of the linear temperature term on doping in the underdoped regimes of hole-doped cuprate, a common experimental observation in different cuprates. It remains puzzling up to now in spite of intensive theoretical efforts. For electron-doped case, even qualitative treatment is not reported at present time. Here we revisit these problems and provide an alternative superfluid density formulation by using the London relation instead of employing the paramagnetic current-current correlation function. The obtained formula, on the one hand, provides the correct temperature-dependent behavior of the superfluid density in the whole temperature regime, on the other hand, makes the doping dependence of the linear temperature term substantially weaken and a possible interpretation for its independence on doping is proposed.
As an application, electron-doped cuprate is studied, whose result qualitatively agrees with existing experiments and successfully explains the origin of $d$- to anisotropic $s$-wave transition across the optimal doping. Our result remedies some failures of the slave-boson theory as employed to calculate superfluid density in cuprates and may be useful in the understanding of the related physics in other strongly correlated systems, e.g. Na$_{x}$CoO$_{2}$$\cdot$yH$_{2}$O and certain iron-based superconductors with dominating local magnetic exchange interaction.
\end{abstract}

\maketitle

\section{Introduction} \label{intr}
The slave-particle theory, which splits the physical electron into various auxiliary bosonic, fermionic and even anyonic elementary degree of freedom, has been successfully utilized in numerous strongly correlated electron systems, ranged from frustrated quantum magnetism,\cite{Sachdev2011,Wen2004} abelian/non-abelian quantum Hall liquid to heavy fermion compounds and high-temperature superconductors.\cite{Halperin1993,Wen2004,Coleman,Senthil2004,Anderson2004,Lee2006}
Among them, the most important achievement of the U(1)/SU(2) slave-boson theory (or the 'plain vanilla' version, the renormalized mean-field theory) is that it has at least qualitatively
captured basic features of the global phase diagram of cuprate superconductor,\cite{Anderson2004,Lee2006} e.g., the underdoped pseudogap behavior, optimally doped strange metal state and overdoped Landau Fermi liquid state.\cite{Timusk1999,Wen1999,Lee2014,Varma1989,Nagaosa1990}

However, in spite of those extreme success, one of the essential physical measurements of any superconductors, i.e. the London penetration depth in Meissner effect, has not yet been correctly reproduced and explained in terms of the mentioned powerful slave-boson technique.\cite{Lee2006} Physically, the temperature-dependent London penetration depth/superfluid density directly detects the superconducting quasiparticle excitation above the highly entangled many-body superconducting ground-state and thus can provide unambiguous clue for the gap structure of superconducting pairing symmetry.\cite{Xiang}

The main challenges for the slave-boson theory in explaining the superfluid density experiments of cuprates are that it cannot reproduce the following two experimental observations, namely, i) the temperature-dependent behavior of superfluid density and ii) doping-independence of the low temperature linear-T term of superfluid density in hole-doped compounds despite of the so-called Uemura scaling (superfluid density is proportional to doping level) has been explained by the theory.\cite{Boyce2000,Stajis2003,Ioffe2002,Lee2006,Yong2012} Due to these failures, it was believed that non-perturbative effects beyond mean-field and Gaussian fluctuations should play an essential role even in the well-formulated superconducting states. More seriously, since almost all calculations in the slave-particle formalism are performed in terms of the framework of mean-field theory and Gaussian gauge fluctuation, these failures may imply the painful breakdown of the slave-particle technique in any unconventional superconductivity.

Furthermore, we notice that in electron-doped cuprate, e.g. Nd$_{2-x}$Ce$_{x}$CuO$_{4}$ and Pr$_{2-x}$Ce$_{x}$CuO$_{4}$, it has been firmly established that both electron and hole Fermi pockets are responsible for the resulting $d$-wave superconducting state.\cite{Luo2005,Das2007,Xiang2009,Armitage2010} However, in the framework of slave boson theory, even qualitative treatment on their London penetration depth is not reported although two Fermi pockets behavior, B$_{1g}$, B$_{2g}$ Raman scattering spectra and inelastic neutron scattering spectra have been well-explained by slave-boson mean-field theory with assumption of $(\pi,\pi)$ anti-ferromagnetic spin-density-wave (SDW) order.\cite{Yuan2004,Liu2006,Liu2007}

But, considering the fact that the slave-boson theory has provided so much interesting physics of cuprates, we hold the point that the conventional (Higgs confined) superconducting state should also be well described by this theory with appropriate extension and/or modification.
In this paper, we focus on the mentioned issue of superfluid density and revisit the formulation of the London penetration depth. Instead of the current-current correlation function, we employ the London relation to calculate the superfluid density. It is found that the obtained formula can reproduce correctly the temperature-dependent behavior of the superfluid density in terms of the usual slave-boson mean-field formalism. Such new formalism emphasizes the weaken quantum correction in the current-current correlation of slave-boson theory accompanying with the disappearance of doping dependence in paramagnetic current response part, thus the result of the low temperature London penetration depth is now consistent with experimental data. More interestingly, at low temperature, the doping dependence of the linear temperature term of superfluid density is found to be substantially weakened and if a physical superconducting gap is considered, independence of doping dependence can be readily realized. [Note, however a fluctuating $d$-wave superconductor model with weakly interacting Bose gas may give rise to the doping-independence.\cite{Herbut2005}]

Moreover, the impurity and multi-band correction are analyzed. It shows weak correlation effect on the dynamics of impurity scattering. The superfluid density formula is derived for the electron-doped cuprate and the calculation is qualitative consistent with experiment. Furthermore, the approximated low temperature formula explains the origin of $d$- to anisotropic $s$-wave transition across the optimal doping, which is a success of the present theory. We hope that the finding obtained in the present work may be also relevant for unconventional superconductivity in triangular lattice compound Na$_{x}$CoO$_{2}$$\cdot$yH$_{2}$O and certain iron-based superconductors with dominating local magnetic exchange interaction.\cite{Takada2003,Si2008}

The remainder of this paper is organized as follows.
In Sec.\ref{sec2}, the mean-field Hamiltonian of usual $t-J$ model on square lattice is introduced and briefly discussed. In Sec.\ref{sec3}, the superfluid density formula in the slave-boson mean-field theory is reviewed. Next, in Sec.\ref{sec4} we provide an alternative formalism for superfluid density, which gives correct result when comparing with experiments. In Sec.\ref{sec5} we give some discussions on the impurity and multi-band effects, which may be important for real materials like electron-doped
cuprate and iron-based superconductors. Finally, we end this work with a brief conclusion in Sec.\ref{sec6}.

\section{The mean-field model}\label{sec2}
The standard mean-field Hamiltonian for the $t-J$-like model in the slave-boson framework reads \cite{Zhang1988,Anderson2004}
\begin{eqnarray}
H&&=\sum_{k\sigma}\varepsilon_{k}f_{k\sigma}^{\dag}f_{k\sigma}+J\sum_{k}\Delta_{k}(f_{k\uparrow}^{\dag}f_{-k\downarrow}^{\dag}+f_{-k\downarrow}f_{k\uparrow}) \nonumber  \\
&&+2J(\chi^{2}+\frac{\Delta^{2}}{2}),\label{eq1}
\end{eqnarray}
where the single-particle energy spectrum including up to the third nearest-neighbor hopping is $\varepsilon_{k}=(-2t\delta-J\chi)(\cos k_{x}+\cos k_{y})-4t'\delta\cos k_{x}\cos k_{y}-4t''\delta(\cos^{2}k_{x}+\cos^{2}k_{y}-1)-\mu$ with $\delta$ denoting the doping level. The strong correlation effect from prohibiting double occupation on one site in the original t-J model is encoded with explicit doping dependence in the energy spectrum.
The mean-field parameters are defined as $\langle f_{i\sigma}^{\dag}f_{j\sigma}\rangle = \chi$ and $\langle f_{i\uparrow}^{\dag}f_{j\downarrow}^{\dag}-f_{i\downarrow}^{\dag}f_{j\uparrow}^{\dag}\rangle = -\Delta_{ij}$, which are from the decoupling of
Heisenberg interaction in the particle-hole and particle-particle channel, respectively.

It is important to recall that the definition of these mean-field parameters is motivated by the local spin-singlet idea of resonance-valence-bond (RVB) quantum liquid\cite{Anderson2004}, thus the local pairing is formed from the beginning and no super-glue like phonon is involved. However, the more conventional spin-fluctuation theory can also give rise similar effective pairing Hamiltonian Eq.\ref{eq1} by extracting singlet paring interaction from the static transverse susceptibility, which
results from the local Heisenberg interaction.\cite{Coleman} Therefore, in this sense, the use of slave-boson mean-field formalism does not necessarily link to RVB but only indicates the strong coupling feature of the problem itself imposed by the non-double occupation condition.

The physical electron is transformed into $c_{\sigma}=\sqrt{\delta}f_{\sigma}$ with $f_{\sigma}$ being the fermionic spinon excitation. It should be emphasized that
although we have used the slave-boson representation to rewrite the electron operator with fractionalized fermionic spinon, the theory itself does not work at deconfined state.
In other words, no fractionalized quantum phase is involved in our calculation and all considered states here are 'normal' phases, which could be understood from general Fermi liquid framework as enthusiastically advocated by Laughlin.\cite{Laughlin2014}

Then, using the Bogoliubov transformation $A_{k\uparrow}=\mu_{k}f_{k\uparrow}+\nu_{k}f_{-k\downarrow}^{\dag},A_{-k\downarrow}^{\dag}=-\nu_{k}f_{k\uparrow}+\mu_{k}f_{-k\downarrow}^{\dag}$ and
$\mu_{k}^{2}=1-\nu_{k}^{2}=\frac{1}{2}\left(1+\frac{\varepsilon_{k}}{E_{k}}\right)$,
we can obtain the diagonalized Hamiltonian as
\begin{eqnarray}
H=&&\sum_{k}E_{k}(A_{k\uparrow}^{\dag}A_{k\uparrow}+A_{-k\downarrow}^{\dag}A_{-k\downarrow})+\sum_{k}(\varepsilon_{k}-E_{k})\nonumber \\
&& +2J(\chi^{2}+\frac{\Delta^{2}}{2})\label{eq2}
\end{eqnarray}
with $E_{k}=\sqrt{\varepsilon_{k}^{2}+J^{2}\Delta_{k}^{2}}$. Usually, on the square lattice, one considers the $d_{x^{2}-y^{2}}$-wave pairing $\Delta_{k}=\Delta(\cos k_{x}-\cos k_{y})$ and the extended $s$-wave pairing $\Delta_{k}=\Delta(\cos k_{x}+\cos k_{y})$. For the cuprate, it has been firmly established that the dominated pairing symmetry is the anisotropic $d_{x^{2}-y^{2}}$-wave,\cite{Lee2006,Armitage2010} although topological $p$+i$p$, $d$+i$d$ and even Fulde-Ferrell-Larkin-Ovchinnikov type pairing structure might be relevant to the heavily underdoped cuprate.\cite{Lu2014,Gupta2014,Das2013} Therefore, we will only consider the case of $d_{x^{2}-y^{2}}$-wave in the remaining part of the present paper, and extension to other pairing symmetry is straightforward.

\section{Superfluid density in slave-boson mean-field theory}\label{sec3}
\begin{figure}[h]
\includegraphics[width=0.8\columnwidth]{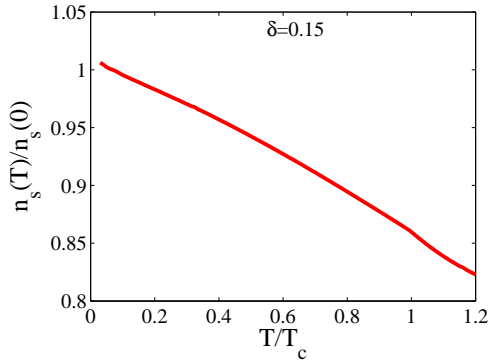}
\caption{\label{fig:1} The failure of the superfluid density formulation Eq. (\ref{eq3}) obtained by the conventional current-current correlation function. The normalized superfluid density $n_{s}(T)/n_{s}(0)$ versus $T/T_{c}$ at doping level $\delta=0.15$ dose not vanish at critical temperature $T_{c}$.}
\end{figure}

The standard superfluid density formula in the framework of slave-boson mean-field theory reads as follows (See Appendix A for details),\cite{Anderson2004}
\begin{eqnarray}
&& \frac{n_{s}}{m}=\delta\sum_{k}\left[\frac{\partial^{2} \varepsilon_{k}}{\partial k_{x}^{2}}\left(1-\frac{\varepsilon_{k}}{E_{k}}\tanh\left(\frac{E_{k}}{2T}\right)\right)\right. \nonumber\\
&&\hskip 2cm \left. +2\delta\left(\frac{\partial \varepsilon_{k}}{\partial k_{x}}\right)^{2}\frac{\partial f_{F}(E_{k})}{\partial E_{k} }\right].\label{eq3}
\end{eqnarray}
Here, the first term, which mainly contributes to the zero-temperature superfluid density, has negligible dependence of temperature. However, in the second term, the thermal excited superconducting quasiparticles deplete the superfluid density,
which leads to the temperature-dependent behavior. It is emphasized that since the second term comes from the paramagnetic current-current correlation, which involves two single-electron Green functions $\langle j^{i}j^{j}\rangle\sim\langle cc^{\dag}\rangle\langle cc^{\dag}\rangle\sim \delta^{2}\langle ff^{\dag}\rangle\langle ff^{\dag}\rangle$, the $\delta^{2}$ dependence is expected. However, as what can be seen in Fig.\ref{fig:1}, the normalized superfluid density $n_{s}(T)/n_{s}(0)$ versus $T/T_{c}$ dose not vanish at critical temperature $T_{c}$, which implies the breakdown of the standard superfluid density formula.

More seriously, if we try to inspect the low temperature behavior of the superfluid density, it reads \cite{Lee2006,Lee1997}
\begin{eqnarray}
n_{s}(T)\approx n_{s}(0)(1-\delta\frac{(2\ln2)T}{\Delta}).\label{eq4}
\end{eqnarray}
with $n_{s}(0)\sim \delta$. It has been emphasized by Lee, Nagaosa and Wen that $n_{s}(0)\sim \delta$ agrees with experiments, however the doping dependence in the linear-$T$ term $\sim\delta^{2}$ dose not.\cite{Lee2006} As observed in existing experimental data, the linear-$T$ term should not show apparent dependence on doping level $\delta$.\cite{Boyce2000,Stajis2003,Ioffe2002} Importantly, such inconsistency leads to the belief that non-perturbative effects beyond mean-field and Gaussian
fluctuation play an essential role even in the well-formulated superconducting states.\cite{Lee2006} Since almost all of calculations in the slave-particle framework are established in terms of the mean-field theory and Gaussian
gauge fluctuation, such failure may break down even the application of slave-particle technique in any unconventional superconductivity.

Theoretically, one expects fluctuation effect beyond mean-field and Gaussian level may cancel out the undesirable doping dependence. In this respect, the numerical variational Monte Carlo method, which exactly performs Gutzwiller projection on each site for mean-field wave-function may give more reliable estimation on corresponding physical quantities but calculation on temperature-dependent superfluid density has still not been reported in literature.\cite{Ogata2008}

\section{Alternative formalism for superfluid density}\label{sec4}
In this section, we will present the details on an alternative formalism for superfluid density, which can overcome the mentioned difficulty. The derivation procedure is conventional in fact and can be found in the standard text books. \cite{Tinkham,Poole} The main point is that the no current-current correlation is used for this derivation, thus the doping effect is weakened. We should remind the reader that similar formalism has been successfully used in the superfluid density calculation about hole/electron-doped cuprate \cite{Das2007,Das2013} and $t-J$ model on the honeycomb lattice.\cite{Zhong2014}

First, we define the so-called London relation (The scalar case is just $\vec{J}=-\frac{n_{s}e^{2}}{m}\vec{A}$.)
\begin{eqnarray}
J_{i}=-Q_{ij}A_{j},\nonumber
\end{eqnarray}
where $Q_{ij}$ is a tensor, which is related to the superfluid density as $Q_{ii}=\frac{n_{s}^{i}e^{2}}{m^{i}}$. And,
$J_{i}$ and $A_{i}$ represent the electronic current density and external electromagnetic vector potential, respectively.
Meanwhile, the current $J_{i}$ has two components, namely, the paramagnetic part $J_{p}^{i}$ and the diamagnetic part $J_{d}^{i}$.

\subsection{Paramagnetic current $J_{p}^{i}$}
By the slave-boson technique, the paramagnetic part $J_{p}^{i}=-nev=-e\sum_{k}(v_{k}^{i}\langle c_{k\uparrow}^{\dag}c_{k\uparrow}\rangle+v_{-k}^{i}\langle c_{-k\downarrow}^{\dag}c_{-k\downarrow}\rangle)$ can be written as
\begin{eqnarray}
J_{p}^{i}&&=-e\delta\sum_{k}v_{k}^{i}(\langle f_{k\uparrow}^{\dag}f_{k\uparrow}\rangle-\langle f_{-k\downarrow}^{\dag}f_{-k\downarrow}\rangle)\nonumber\\
&&=-e\delta\sum_{k}v_{k}^{i}(\langle A_{k\uparrow}^{\dag}A_{k\uparrow}\rangle-\langle A_{-k\downarrow}^{\dag}A_{-k\downarrow}\rangle)\nonumber\\
&&=-e\delta\sum_{k}v_{k}^{i}(f_{F}(E_{k})-f_{F}(E_{-k}))\nonumber\\
&&=-e\delta\sum_{k}v_{k}^{i}\left(-2e\frac{\partial f_{F}(E_{k})}{\partial E_{k}}v_{k}^{j}A_{j}\right)\nonumber\\
&&=2e^{2}\delta\sum_{k}\left(v_{k}^{i}v_{k}^{j}\frac{\partial f_{F}(E_{k})}{\partial E_{k}}\right)A_{j}\nonumber.
\end{eqnarray}
In the above derivation, we have used the relation that when there is the external vector potential $\vec{A}$, $f_{F}(E_{k})-f_{F}(E_{-k})\simeq -2e\frac{\partial f_{F}(E_{k})}{\partial E_{k}}v_{k}^{j}A_{j}$. Thus, we have $Q_{ij}^{p}=2e^{2}\delta\sum_{k}[v_{k}^{i}v_{k}^{j}\frac{\partial f_{F}(E_{k})}{\partial E_{k}}]$. (The effective velocity is defined by $v_{k}^{i}=\frac{\partial \varepsilon_{k}}{\partial k_{i}}$.)
It is clear to see that in contrast to the conventional formula in last section, the dependence on doping level is $\sim\delta$ rather than $\delta^{2}$, thus the encountered doping dependence problem actually disappears.
The reason of this distinction is that here the contribution of the paramagnetic current to the superfluid response is effectively single-particle-like rather than the two-particle correlation in the usual current-current correlation function. Usually, as seen from the more sophisticated variational Monte Carlo calculation,\cite{Li2011} the slave-boson mean-field theory often overestimates the effect of strongly electronic correlations and the result presented here suppresses partial unphysical degree of freedom introduced in slave-particle representation.

\subsection{Diamagnetic current $J_{d}^{i}$}
Correspondingly, the diamagnetic part $J_{d}^{i} = -e^{2}\frac{1}{m_{ij}}A_{j}n = -e^{2}A_{j}\sum_{k}\frac{1}{m_{k}^{ij}}(\langle c_{k\uparrow}^{\dag}c_{k\uparrow}\rangle+\langle c_{-k\downarrow}^{\dag}c_{-k\downarrow}\rangle) $ ($\frac{1}{m_{k}^{ij}}=\frac{\partial^2 \varepsilon_{k}}{\partial k_{i}\partial k_{j}}$) can be written as
\begin{eqnarray}
J_{d}^{i}&&=-e^{2}\delta A_{j}\sum_{k}\frac{1}{m_{k}^{ij}}\left(\langle f_{k\uparrow}^{\dag}f_{k\uparrow}\rangle+\langle f_{-k\downarrow}^{\dag}f_{-k\downarrow}\rangle\right)\nonumber\\
&&=-e^{2}\delta A_{j}\sum_{k}\left[\frac{1}{m_{k}^{ij}}(\mu_{k}^{2}-\nu_{k}^{2})(\langle A_{k\uparrow}^{\dag}A_{k\uparrow}\rangle+\langle A_{-k\downarrow}^{\dag}A_{-k\downarrow}\rangle) \right. \nonumber\\
&& \hskip 2cm \left.+\frac{1}{m_{k}^{ij}}2\nu_{k}^{2}\right]\nonumber\\
&&=-e^{2}\delta A_{j}\sum_{k}\left[\frac{1}{m_{k}^{ij}}\frac{\varepsilon_{k}}{E_{k}}2f_{F}(E_{k})+\frac{1}{m_{k}^{ij}}(1-\frac{\varepsilon_{k}}{E_{k}})\right]\nonumber\\
&&=-e^{2}\delta A_{j}\sum_{k}\left[\frac{1}{m_{k}^{ij}}\left(1-\frac{\varepsilon_{k}}{E_{k}}\tanh(\frac{E_{k}}{2T})\right)\right].\nonumber
\end{eqnarray}
Therefore, the diamagnetic kernel reads $Q_{ij}^{d}=e^{2}\delta\sum_{k}\left[\frac{1}{m_{k}^{ij}}\left(1-\frac{\xi_{k}}{E_{k}}\tanh(\frac{E_{k}}{2T})\right)\right]$.

\subsection{The superfluid density $n_{s}$}
\begin{figure}
\includegraphics[width=0.8\columnwidth]{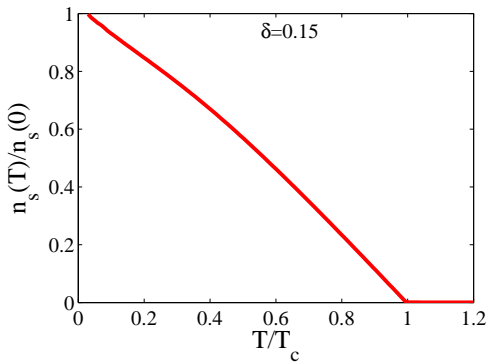}
\caption{\label{fig:2} The qualitatively correct temperature-dependence of the superfluid density given by Eq. (\ref{eq5}). The normalized superfluid density $n_{s}(T)/n_{s}(0)$ versus $T/T_{c}$ at doping level $\delta=0.15$ by using obtained superfluid density formalism.}
\end{figure}

\begin{figure}[tbp]
\centering
\includegraphics[width = 0.45\columnwidth]{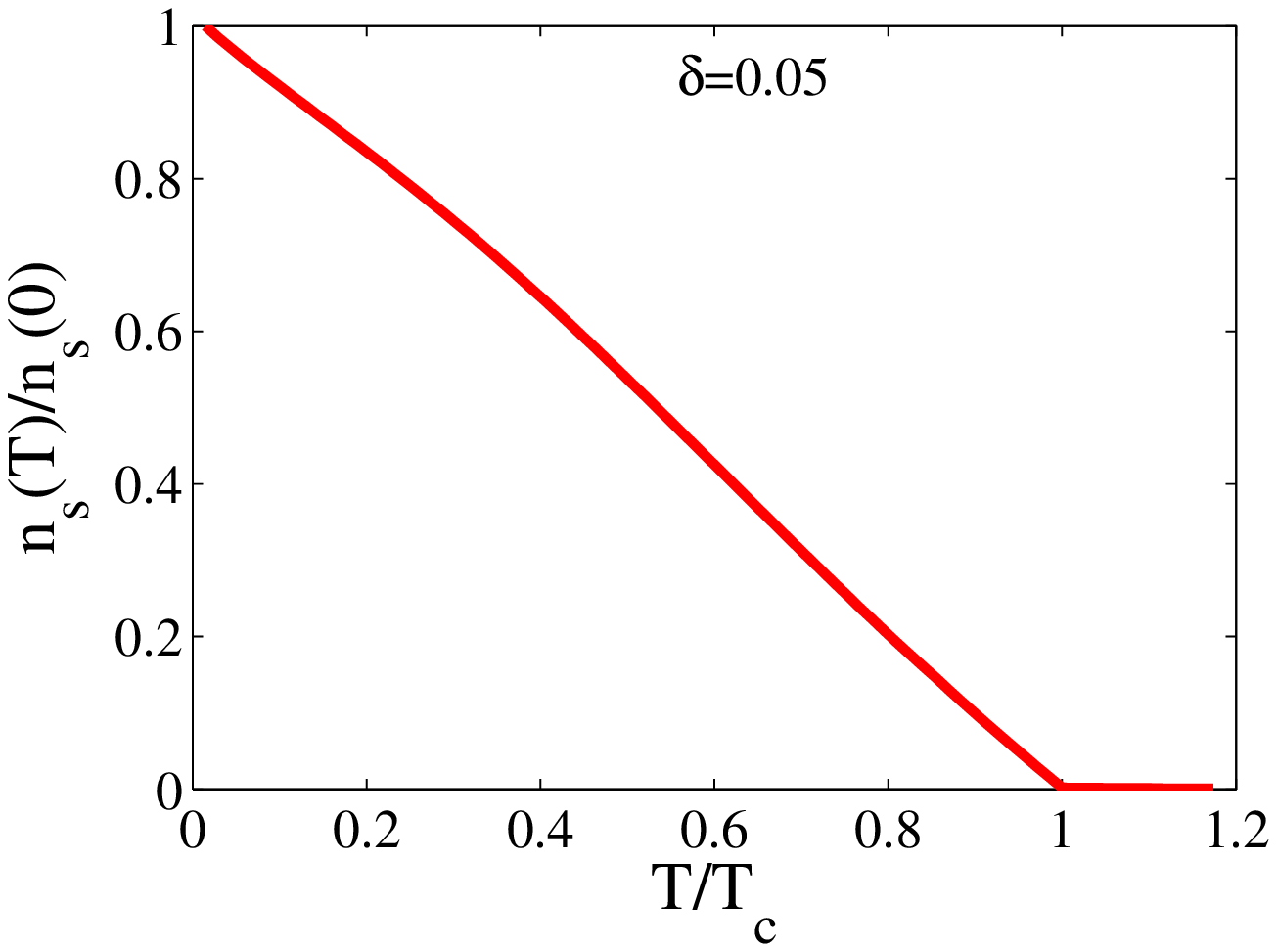}
\includegraphics[width = 0.45\columnwidth]{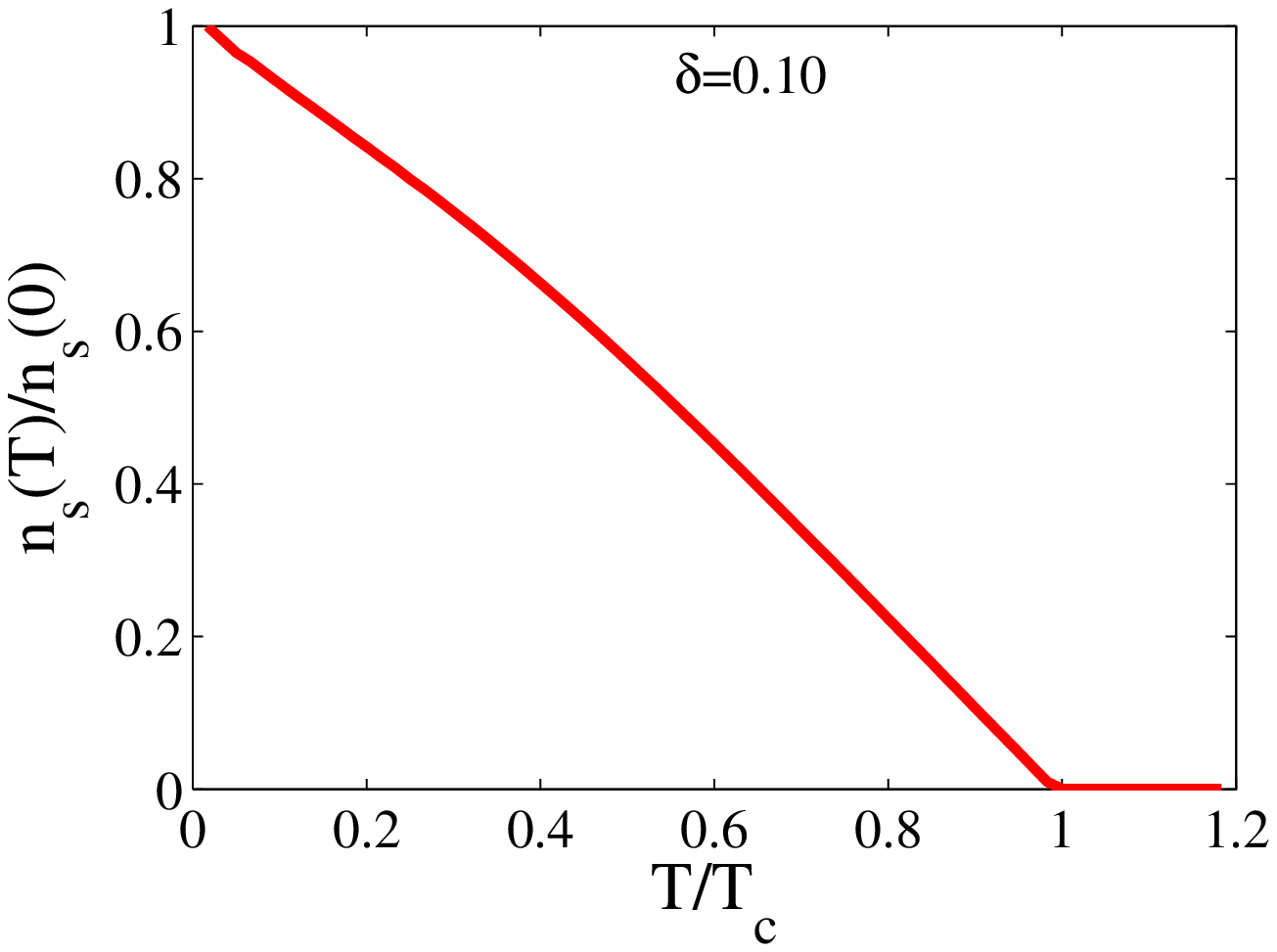}
\includegraphics[width = 0.45\columnwidth]{SF2}
\includegraphics[width = 0.45\columnwidth]{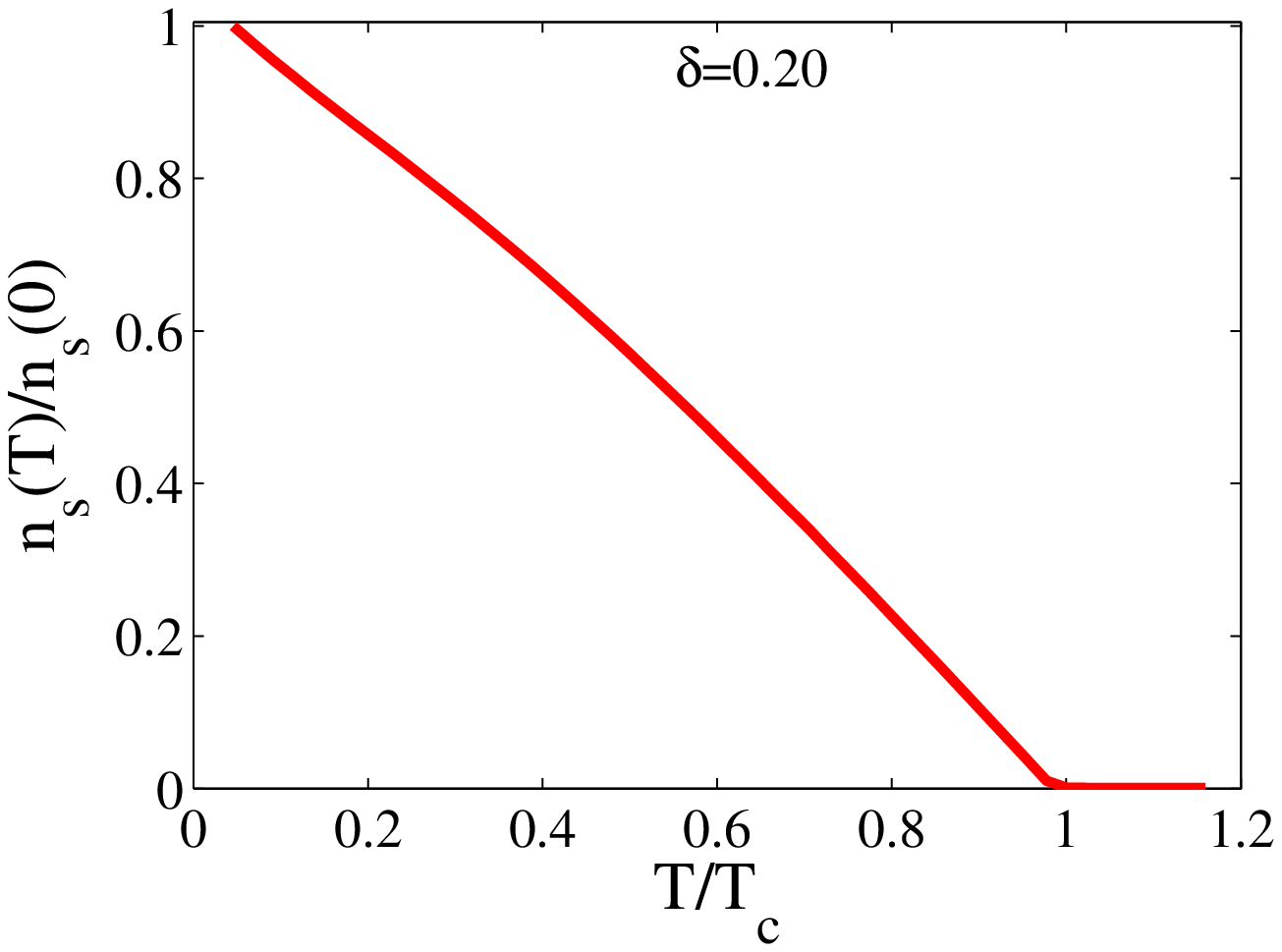}
\caption{\label{fig3} The normalized superfluid density $n_{s}(T)/n_{s}(0)$ at different doping levels ranged from underdoped to overdoped regimes.}
\end{figure}

Since the total superfluid response kernel $Q_{ij}=Q_{ij}^{p}+Q_{ij}^{d}$, we have
\begin{equation}
Q_{ij}=e^{2}\delta\sum_{k}\left[\frac{1}{m_{k}^{ij}}(1-\frac{\varepsilon_{k}}{E_{k}}\tanh(\frac{E_{k}}{2T}))+2v_{k}^{i}v_{k}^{j}\frac{\partial f_{F}(E_{k})}{\partial E_{k}}\right] \nonumber
\end{equation}
and
\begin{equation}
Q_{ii}=e^{2}\delta\sum_{k}\left[\frac{1}{m_{k}^{ii}}(1-\frac{\varepsilon_{k}}{E_{k}}\tanh(\frac{E_{k}}{2T}))+2(v_{k}^{i})^{2}\frac{\partial f_{F}(E_{k})}{\partial E_{k}}\right].\nonumber
\end{equation}
So, the superfluid density reads
\begin{eqnarray}
\frac{n_{s}^{i}}{m^{i}}&&=\delta\sum_{k}\left[\frac{1}{m_{k}^{ii}}(1-\frac{\varepsilon_{k}}{E_{k}}\tanh(\frac{E_{k}}{2T}))+2(v_{k}^{i})^{2}\frac{\partial f_{F}(E_{k})}{\partial E_{k}}\right]\nonumber\\
&&=\delta\sum_{k}\left[\frac{\partial^{2} \varepsilon_{k}}{\partial k_{i}^{2}}(1-\frac{\varepsilon_{k}}{E_{k}}\tanh(\frac{E_{k}}{2T}))+2(\frac{\partial \varepsilon_{k}}{\partial k_{i}})^{2}\frac{\partial f_{F}(E_{k})}{\partial E_{k}}\right]\nonumber.
\end{eqnarray}
Usually, since the considered system is symmetric between $x$ and $y$ direction, which appears in the slave-boson mean-field theory on square lattice, we can use the simplified formula below
\begin{equation}
\frac{n_{s}}{m}=\delta\sum_{k}\left[\frac{\partial^{2} \varepsilon_{k}}{\partial k_{i}^{2}}\left(1-\frac{\varepsilon_{k}}{E_{k}}\tanh(\frac{E_{k}}{2T})\right)+2(\frac{\partial \varepsilon_{k}}{\partial k_{i}})^{2}\frac{\partial f_{F}(E_{k})}{\partial E_{k}}\right].\label{eq5}
\end{equation}
Obviously, one can observe that the undesirable doping dependence does not appear in the quasiparticles depletion term (the second term), which leads to the physically correct result as shown in Fig. \ref{fig:2} and qualitatively agrees with
the penetration depth measurement in hole-doped cuprate superconductor.\cite{Hardy1993,Panagopoulos1998,Yong2012}
Furthermore, the formulation obtained can be employed to all doping cases, ranged from underdoped to overdoped regime, as shown in Fig. \ref{fig3}. [Obviously, since no peusdo-gap is introduced in the present model, the comparison to underdoped cuprate should not be considered seriously. It seems that the phenomenological theory in Ref.\onlinecite{Rice2012} may be useful in treating the issue of peusdo-gap.] The result is also consistent with reported data in Refs. [\onlinecite{Boyce2000}] and [\onlinecite{Panagopoulos1998}]. In other words, the curvature of the normalized superfluid density $\frac{d}{dT} \left(\frac{n_{s}(T)}{n_{s}(0)}\right)$ is nearly identical and is independent of doping level, thus recovers the observed behavior in experiments.

Consider the low temperature behavior of the superfluid density, we have
\begin{equation}
n_{s}(T)\approx n_{s}(0)(1-\frac{(2\ln2)T}{\Delta}).\label{eq6}
\end{equation}
Comparing with Eq. (\ref{eq4}), the doping dependence of linear term is weakened to $\sim\delta$ instead of $\sim\delta^{2}$, which improves significantly the slave-boson theory at low temperature. However, both the experimental data and phenomenological theoretical analysis of superfluid density suggest
the following low temperature formula,\cite{Lee1997,Boyce2000,Stajis2003,Ioffe2002,Yong2012}
\begin{equation}
n_{s}(T)-n_{s}(0)\sim-\frac{T}{\Delta(0)},\label{eq7}
\end{equation}
which shows no apparent dependence of doping level. (The gap $\Delta(0)$ may only have weak dependence of doping.)
If one considers the fact that in the slave-boson theory, the true superconducting gap
of original electron is $\Delta_{sc}=\delta\Delta$, one has
\begin{equation}
n_{s}(T)-n_{s}(0)\approx-n_{s}(0)\frac{(2\ln2)T}{\Delta_{sc}}\sim -\frac{T}{\Delta}.\label{eq8}
\end{equation}
This formula is identical to Eq. (\ref{eq7}). Therefore, an improved slave-boson theory description with the physical pairing gap is able to reconcile the difficulty between the present theory and the experimental measurements.

\section{Impurity and multi-band effect}\label{sec5}
For realistic materials, the nonmagnetic impurity scattering is ubiquitous and plays a crucial role in the low temperature behaviors of any superconductors, particularly for nodal pairing states like the most important d-wave state. On the other hand, in many real-life materials, many energy bands contribute to the ultimate superconductivity, thus it is also helpful to investigate the formalism of superfluid density in the multi-band case.

\subsection{Non-magnetic impurity}
Following Ref. [\onlinecite{Goldenfeld1993}], when considering the effect of nonmagnetic impurity on the nodal d-wave state, the corresponding
low temperature superfluid density reads
\begin{equation}
\tilde{n}_{s}(T)\approx \tilde{n}_{s}(0)\left(1-2\frac{T^2}{T+T^{\ast}}\right),\label{eq9}
\end{equation}
where the zero-temperature superfluid density $\tilde{n}_{s}(0)\approx n_{s}(0)/(1+1.58\sqrt{\Gamma/\Delta})$ is suppressed by the impurity scattering $\Gamma$ and there is a crossover temperature $T^{\ast}\approx0.83\sqrt{\Gamma\Delta}$.
Furthermore, if $T\ll T^{\ast}$, the $T^{2}$ behavior can be seen in the superfluid density instead of the linear behavior at elevated $T\gg T^{\ast}$.
It is noted that the only effect of strong electron correlation is $n_{s}(0)\approx\delta$. This is because that the dynamics of impurity scattering is fully encoded by the auxiliary fermions without involving any real charge fluctuation.

\subsection{Multi-band effect: Application to electron-doped cuprate}
Generally, the hole-doped cuprate permits a single-band description and the formalism presented in the main text is applicable.

For the case of electron-doped cuprate, e.g. Nd$_{2-x}$Ce$_{x}$CuO$_{4}$ and Pr$_{2-x}$Ce$_{x}$CuO$_{4}$, it has been firmly established that both electron and hole Fermi pockets are responsible for the resulting $d$-wave superconducting state.\cite{Armitage2010}
Theoretically, the $t-J$ model has been utilized with inverse sign of hopping parameters as compared to their hole-doped counterpart.\cite{Tohyama2001} The observed two Fermi pockets behavior is captured by the slave-boson mean-field theory with with assumption of $(\pi,\pi)$ anti-ferromagnetic spin-density-wave (SDW) order.\cite{Yuan2004} Other physical quantities like B$_{1g}$, B$_{2g}$ Raman scattering spectra and inelastic neutron scattering spectra have also been well-explained by slave boson theory.\cite{Liu2006,Liu2007} However, we emphasize that there is still no qualitative treatment on the London penetration depth in those electron-doped cuprate. [Note, however a phenomenological two-band model has succeed in fitting and explaining on the experimental data by our previous work.\cite{Luo2005}]

\begin{figure}[tbp]
\centering
\includegraphics[width = 0.8\columnwidth]{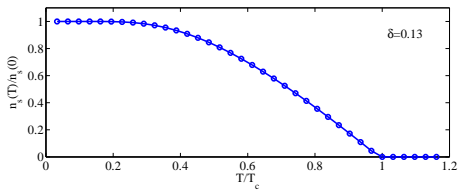}
\caption{\label{fig4} The normalized superfluid density $n_{s}(T)/n_{s}(0)$ at optimal doping levels $\delta=0.13$ for electron-doped case.}
\end{figure}

Here, utilize the $t-J$-like model in previous works and extend the discussion in last section,\cite{Yuan2004,Liu2006,Liu2007} we can obtain the following superfluid density formula,
\begin{equation}
\frac{n_{s}}{m}=\delta\sum_{k,\alpha=\pm}\left[-\frac{\partial^{2} \xi_{k}^{\alpha}}{\partial k_{x}^{2}}\frac{\xi_{k}^{\alpha}}{E_{k}^{\alpha}}\tanh\frac{E_{k}^{\alpha}}{2T}+2(\frac{\partial \xi_{k}^{\alpha}}{\partial k_{x}})^{2}\frac{\partial f_{F}(E_{k}^{\alpha})}{\partial E_{k}^{\alpha}}\right].\nonumber
\end{equation}
Here, $\xi_{k}^{\pm}=(\varepsilon_{k}+\varepsilon_{k+Q}\pm\sqrt{(\varepsilon_{k}-\varepsilon_{k+Q})^{2}+(Jm)^{2}})/2$ denotes two antiferromagnetic energy bands, which results from the preformed
anti-ferromagnetic SDW long-ranged order with $Q=(\pi,\pi)$.\cite{Yuan2004} If no doping is introduced , the system is an anti-ferromagnetic Mott insulator. Upon doping, the $\xi_{k}^{+}$ band gives rise to the electron-like Fermi surface centered at $(\pi,0)$. When approaching optimal doping level, the $\xi_{k}^{-}$ band drives the formation of hole-like Fermi surface around ($\pi/2,\pi/2$) and the superconducting instability ultimately develops with appearance of such hole Fermi surface. The corresponding superconducting quasi-particle spectrum is $E_{k}^{\alpha}=\sqrt{(\xi_{k}^{\alpha})^{2}+\Delta_{k}^{2}}$ with the assumption of $d_{x^{2}-y^{2}}$-wave gap $\Delta_{k}$.\cite{Liu2006} In Fig.\ref{fig4}, we have shown the normalized superfluid density at doping levels $\delta=0.13$ (optimal doped), which is consistent with the published data of the penetration depth measurement in Ref.\onlinecite{Kim2003}.

More importantly, when inspecting the low temperature behavior of the superfluid density, we find
\begin{equation}
n_{s}(T)=n_{s}^{+}(0)(1-a\sqrt{\frac{\Delta'}{T}})e^{-\Delta'/T}+n_{s}^{-}(0)(1-b\frac{T}{T_{c}})                  .\nonumber
\end{equation}
Here, $\Delta'$ is the minimum value of gap function $\Delta_{k}$ at the $(\pi,0)$ electron Fermi surface with constant $a$ and $b$. With this approximated formula, one can see that when doping is not large ($n_{s}^{+}(0)\gg n_{s}^{-}(0)$), only the first term dominates, which leads to the gapped $s$-like exponential behavior observed in experiments.\cite{Kim2003} When the hole Fermi surface forms around optimal doping, the latter term with linear T behavior competes with the first term, thus behaves like a gapless nodal superconductor. Therefore, this simplified formalism, which agrees well with the results in our previous study,\cite{Luo2005} suggests that there is no true $d$- to anisotropic $s$-wave transition across the optimal doping.\cite{Kim2003}

\subsection{Multi-band effect: Iron-based superconductors}
When entering the `iron age', it has been suggested that certain iron-based superconductors can be viewed from the perspective of the doped Mott or orbital selective Mott insulator, whose pairing originates from the local magnetic exchange interaction in the parent FeAs and FeTe compounds.\cite{Si2008,Hu2012} Thus, their pairing physics may be explored by multi-band version of $t-J$ model since more than one Fe $d$-orbital contribute to the low energy physics. Particularly, the superfluid density formula derived in the main text can be used in this case with proper extension to include orbital effect.

\section{Conclusion and Discussion}\label{sec6}
We have provided an alternative formalism for superfluid density, which can reproduce qualitatively the existing experimental data in high-T$_{c}$ hole and electron-doped cuprate superconductors. This new formalism resolves the failure of slave-boson theory as applied to calculate the temperature-dependent superfluid density in $t-J$-like strong coupling lattice fermion models. It also indicates that the slave-boson mean-field theory generically overestimates the many-particle correlation effect in the many-point correlation function, which should be further studied by more sophisticated techniques like variational Monte Carlo simulation and cluster dynamic mean-field theory. Therefore, although only superfluid density is analyzed in the present work, we believe that other physical observable dominated by many-particle correlation, e.g. optical conductance and Hall coefficient, should be calculated with similar formalism as ours when slave-boson theory is applied.

\begin{acknowledgments}
We thank Jianhui Dai and Tao Li for helpful discussion on the proposed superfluid density formalism. The work is partly supported by the programs for NSFC, PCSIRT (Grant No. IRT1251), the national program for basic research and the Fundamental Research Funds for the Central Universities of China.
\end{acknowledgments}

\appendix
\section{The standard superfluid density formula in slave-boson mean-field theory}
The standard superfluid density $n_{s}^{i}$ formula reads as follows \cite{Xiang}
\begin{eqnarray}
\frac{n_{s}^{i}}{m^{i}}=\sum_{k\sigma}\frac{\partial^{2} \varepsilon_{k}}{\partial k_{i}^{2}}\langle c_{k\sigma}^{\dag}c_{k\sigma}\rangle-\int d\tau\langle j^{i}(0)j^{i}(\tau)\rangle
\end{eqnarray}
where $\langle \cdot \rangle$ means the expectation in the BCS mean-field Hamiltonian and
\begin{eqnarray}
j^{i}(\tau)&&=e\sum_{k\sigma}\frac{\partial \varepsilon_{k}}{\partial k_{i}}c_{k\sigma}^{\dag}(\tau)c_{k\sigma}(\tau) \nonumber \\
&&=e\delta\sum_{k}\frac{\partial \varepsilon_{k}}{\partial k_{i}}f_{k\sigma}^{\dag}(\tau)f_{k\sigma}(\tau) \nonumber \\
&&=\delta j_{f}^{i}(\tau).
\end{eqnarray}
Thus, we have
\begin{equation}
\frac{n_{s}^{i}}{m^{i}}=\delta m^{i}\sum_{k\sigma}\frac{\partial^{2} \varepsilon_{k}}{\partial k_{i}^{2}}\langle f_{k\sigma}^{\dag}f_{k\sigma}\rangle-\delta^{2}\int d\tau\langle j_{f}^{i}(0)j_{f}^{i}(\tau)\rangle.
\end{equation}
After calculating the expectation in the above equation, the final result reads
\begin{equation}
\frac{n_{s}^{i}}{m^{i}}=\delta\sum_{k}\left[\frac{\partial^{2} \varepsilon_{k}}{\partial k_{i}^{2}}(1-\frac{\varepsilon_{k}}{E_{k}}\tanh(\frac{E_{k}}{2T}))+2\delta(\frac{\partial \varepsilon_{k}}{\partial k_{i}})^{2}\frac{\partial f_{F}(E_{k})}{\partial E_{k} }\right].\nonumber
\end{equation}

\end{document}